\documentclass[10pt]{bmc_article}

\usepackage{graphicx}
\usepackage{cite} % Make references as [1-4], not [1,2,3,4]
\usepackage{url}  % Formatting web addresses  
\usepackage{ifthen}  % Conditional 
\usepackage{multicol}   %Columns
\usepackage[utf8]{inputenc} %unicode support
\newboolean{publ}

\newenvironment{bmcformat}{\fussy\setboolean{publ}{true}}{\fussy}

\begin{document}
\begin{bmcformat}


%\title{Recognition of protein folds using amino acid occurrence}
\title{Comparison of amino acid occurrence and composition for predicting 
protein folds}
 
%%%%%%%%%%%%%%%%%%%%%%%%%%%%%%%%%%%%%%%%%%%%%%
%%                                          %%
%% Enter the authors here                   %%
%%                                          %%
%% Ensure \and is entered between all but   %%
%% the last two authors. This will be       %%
%% replaced by a comma in the final article %%
%%                                          %%
%% Ensure there are no trailing spaces at   %% 
%% the ends of the lines                    %%     	
%%                                          %%
%%%%%%%%%%%%%%%%%%%%%%%%%%%%%%%%%%%%%%%%%%%%%%

\author{Y-h. Taguchi\correspondingauthor$^{1,2}$%
       \email{Y-h. Taguchi\correspondingauthor - tag@granular.com}%
       and
         M. Michael Gromiha\correspondingauthor$^3$%
         \email{M. Michael Gromiha\correspondingauthor - michael-gromiha@aist.go.jp}
      }

%%%%%%%%%%%%%%%%%%%%%%%%%%%%%%%%%%%%%%%%%%%%%%
%%                                          %%
%% Enter the authors' addresses here        %%
%%                                          %%
%%%%%%%%%%%%%%%%%%%%%%%%%%%%%%%%%%%%%%%%%%%%%%

\address{%
    \iid(1)Department of Physics, Faculty of Science and
Technology, Chuo University, 1-13-27 Kasuga,
Bunkyo-ku, Tokyo
112-8551, Japan.\\
    \iid(2)Institute for
Science and Technology, Chuo University, 1-13-27 Kasuga,
Bunkyo-ku, Tokyo
112-8551, Japan.\\
    \iid(3)Computational Biology Research Center (CBRC), National Institute of Advanced Industrial Science and Technology (AIST), AIST Tokyo Waterfront Bio-IT Research Building, 2-42 Aomi, Koto-ku, Tokyo 135-0064, Japan}%

\maketitle

%%%%%%%%%%%%%%%%%%%%%%%%%%%%%%%%%%%%%%%%%%%%%%
%%                                          %%
%% The Abstract begins here                 %%
%%                                          %%
%% The Section headings here are those for  %%
%% a Research article submitted to a        %%
%% BMC-Series journal.                      %%  
%%                                          %%
%% If your article is not of this type,     %%
%% then refer to the Instructions for       %%
%% authors on http://www.biomedcentral.com  %%
%% and change the section headings          %%
%% accordingly.                             %%   
%%                                          %%
%%%%%%%%%%%%%%%%%%%%%%%%%%%%%%%%%%%%%%%%%%%%%%

\begin{abstract}
        % Do not use inserted blank lines (ie \\) until main body of text.
        \paragraph*{Background:} Prediction of protein three-dimensional structures from amino acid sequences is a long-standing goal in computational/molecular biology. The successful discrimination of protein folds would help to improve the accuracy of protein 3D structure prediction. 
      
        \paragraph*{Results:} In this work, we propose a method based on
linear discriminant analysis (LDA) for recognizing proteins belonging to 30 different folds using the occurrence of amino acid residues in a set of 1612 proteins. The present method could discriminate the globular proteins from 30 major folding types with the sensitivity of 37\%, which is comparable to or better than other methods in the literature. 
A web server has been developed for predicting the folding type of the protein from amino acid sequence and it is available at http://granular.com/PROLDA/.

        \paragraph*{Conclusions:} Linear discriminant analysis based on amino
acid occurrence could successfully recognize protein folds. The present method
has several advantages such as, 
(i) it directly predicts the folding type of a protein without
performing pair-wise comparisons, (ii) it can discriminate folds among large
number of proteins and (iii) it is very fast to obtain the results. This is a simple method, which can be easily incorporated in any other structure prediction algorithms. 
\end{abstract}


\ifthenelse{\boolean{publ}}{\begin{multicols}{2}}{}

%%%%%%%%%%%%%%%%%%%%%%%%%%%%%%%%%%%%%%%%%%%%%%
%%                                          %%
%% The Main Body begins here                %%
%%                                          %%
%% The Section headings here are those for  %%
%% a Research article submitted to a        %%
%% BMC-Series journal.                      %%  
%%                                          %%
%% If your article is not of this type,     %%
%% then refer to the instructions for       %%
%% authors on:                              %%
%% http://www.biomedcentral.com/info/authors%%
%% and change the section headings          %%
%% accordingly.                             %% 
%%                                          %%
%% See the Results and Discussion section   %%
%% for details on how to create sub-sections%%
%%                                          %%
%% use \cite{...} to cite references        %%
%%  \cite{koon} and                         %%
%%  \cite{oreg,khar,zvai,xjon,schn,pond}    %%
%%  \nocite{smith,marg,hunn,advi,koha,mouse}%%
%%                                          %%
%%%%%%%%%%%%%%%%%%%%%%%%%%%%%%%%%%%%%%%%%%%%%%

%%%%%%%%%%%%%%%%
%% Background %%
%%
\section*{Background}

 Deciphering the native conformation of a protein from its amino acid sequence
known as, protein folding problem is a challenging task. The recognition of
proteins of similar folds and/or proteins belonging to same structural class
is a key intermediate step for protein structure prediction. For the past
several decades several methods have been proposed for predicting protein
structural classes. These methods include discriminant analysis\cite{Klein},
correlation coefficient\cite{Chou_2}, hydrophobicity profiles\cite{Gromiha95},
amino acid index \cite{Bu}, Bayes decision rule\cite{Wang}, amino
acid distributions\cite{Kumarevel}, functional domain occurrences \cite{Cai}, supervised fuzzy
clustering approach\cite{Shen05}, amino acid principal component
analysis\cite{Du} etc. These methods showed that the sensitivity lies in the
range of 70-100\% for discriminating protein structural classes and the
sensitivity mainly depends on the dataset. Wang and Yuan\cite{Wang} developed a dataset of 674 globular protein domains belonging to different structural classes and reported that methods claiming 100\% sensitivity for structural class prediction, predicted only with the sensitivity of 60\% with this dataset.

	On the other hand, alignment profiles have been widely used for
recognizing protein folds\cite{Shi,Zhou}. Recently,
Cheng and Baldi\cite{Cheng} proposed a machine learning algorithm for fold
recognition using secondary structure, solvent accessibility, contact map and
$\beta$-strand pairing, which showed the pairwise sensitivity of 27\%. On the
other hand, it has been reported that the amino acid properties are the key
determinants of protein folding and are used for discriminating membrane
proteins\cite{Gromiha05}, identification of membrane spanning
regions\cite{Hirokawa}, prediction of protein structural classes\cite{Chou}, protein folding rates
\cite{Gromiha06}, protein stability\cite{Gromiha99} etc. Towards this direction,
Ding and Dubchak\cite{Ding} proposed a method based on neural networks and support vector machines for fold recognition using amino acid composition  and five other properties, and reported a cross-validated sensitivity of 45 \%. 

 In this work, we have used the amino acid occurrence (not composition) of proteins belonging to 30 major folds  for recognizing protein folds. We have developed a method based on linear discriminant analysis (LDA), which showed an accuracy of 37\% in recognizing 1612 proteins from 30 different folds, which is comparable with  other methods in the literature, in spite of the simplicity of our method and the large number of proteins considered.

%%%%%%%%%%%%%%%%%%%%%%%%%%%%
%% Results and Discussion %%
%%
\section*{Results and Discussion}
    \subsection*{Role of re-weighting for fold recognition}
 We have computed the occurrence of all the 20 amino acid residues in each protein. 
The occurrence of 20 types of residues represents the elements of 20 dimensional vectors for each protein. We have applied LDA to these vectors for recognition. 
Here, we have employed two kinds of LDA, i.e., with and without reweighing.
In LDA with re-weighting, i.e. $W_k=1$ in eq. (\ref{eq:SB}), each fold equally contributes to the measure of performance irrespective of the number of proteins in each fold; i.e., even if one fold includes hundreds of proteins and another has only few proteins, LDA is optimized  to achieve the highest performance equally in each fold. This re-weighting is important especially when the number of proteins included in each fold has large variations.

 On the other hand, LDA without re-weighting, i.e. $W_k=N_k$ in eq. (\ref{eq:SB}), tends to achieve the maximum sensitivity for the whole dataset. In this case, folds with less number of proteins have a strong tendency to be ignored. In accordance with this choice, we have defined two kinds of sensitivities, (i) averaged over folding types and (ii) overall. The overall sensitivity is the ratio between the number of correctly predicted proteins (true positives) in each fold and total number of proteins. Folding type sensitivity is computed as the average of sensitivities obtained in each fold.

 In Table 
%\ref{table:sens}
1, we presented two types of sensitivities (overall and fold average) with two kinds of LDA (with and without re-weighting). 
We observed that re-weighting significantly changed the performance. This is due to the divergence in the number of proteins in each fold (min. 25, max. 173, mean 54,
see Table %\ref{table:sens. set 1}
2). Two kinds of sensitivities differ from each other by almost 10\%, without re-weighting. We achieved the sensitivity of 37\%, which is the best performance to our knowledge, for large number of folds (30) and proteins (1612) considered.  
Further, the method is extremely simple, which indicates that the physical properties of proteins carry sufficient information instead of sequences.

\subsection*{ Prediction of proteins belonging to different folding types}
We have examined the ability of the present method for predicting proteins belonging to 30 major folds. 
In Table %\ref{table:sens. set 1}
2, we have shown the sensitivity of recognizing 30 different folds. We observed that the sensitivity of folds with fewer proteins has increased after re-weighting. All the folds  that have the sensitivity of less than 10 \% without re-weighting are the ones with fewer proteins. For example, 
SAM domain like fold has the sensitivity of 7\%, which has only 26 proteins. Similar tendency is also observed for the folds b.2, b.34, c.3, c.47, c.55, d.15 and d.17.
  On the other hand, many folds with less than 30 proteins have the sensitivity of more than 20\% even without re-weighting (e.g., a.3, a.24, a.39 etc.). As there are 30 folds, the expected sensitivity is only 3.3 \% if classification is supposed to be random. Hence the sensitivity of 20\% obtained for several folds is significantly higher than that of random for fold recognition. Interestingly most of the folds, which have more than 20\% sensitivity, in spite of less number of proteins, belong to either
all-$\alpha$ or all-$\beta$. This might be due to the fact that the proteins belonging to 
all-$\alpha$ and all-$\beta$ classes have different secondary structural patterns and hence they are easy to discriminate them. In addition, folds in these classes are near-by each other in amino acid occurrence vector space, which caused high sensitivity.
The comparison between experimental vs predicted folds is shown in Fig.
%\ref{fig:matrix}
1. In this figure, dark block indicates the presence of relatively higher number of proteins and the data are normalized so that the total percentage of true fold is 100 \%.
We noticed that before re-weighting (Fig. %\ref{fig:matrix}
1(a)), the folds, 
to which many proteins are misclassified, are the ones with more number of proteins (e.g., a.4, b.1, c.1 and d.58). On the other hand, after re-weighting, the trend has been changed: the misclassified proteins mainly accommodates within the same structural class. 
Especially, in $\alpha + \beta$, the block diagonal region is filled almost uniformly, which is partially caused by re-weighting. Since each fold is equally weighted, $\alpha+\beta$  class is less weighted than other classes. 
This causes inter-class misclassification between
$\alpha+\beta$ and other classes, because $\alpha+\beta$ class includes only three folds. 
This problem can be clearly seen in Table %\ref{table:sens. cross. h}
3(a) where we have shown true vs predicted classes with re-weighting. Here, the  classes are not evenly sampled and $\alpha/\beta$ class keeps almost three times as large as 
$\alpha+\beta$. Further, 
neither all-$\alpha$ nor all-$\beta$ are mainly misclassified into
$\alpha+\beta$. 

\subsection*{Hierarchical re-weighting}
In order to resolve this problem, we proposed the scheme of hierarchical re-weighting. In this method, weight is equally distributed to 5
classes (all-$\alpha$, all-$\beta$, $\alpha/\beta$, $\alpha+\beta$, and
small) then it is re-distributed to each fold.
For example, fold  c.37  gets 0.02 weight since it gets one tenth of weight
0.2, which is delivered to $\alpha/\beta$.
The results obtained with hierarchical re-weighting is also included in Table 
%\ref{table:sens. set 1}
2.
The comparison of results obtained with and without re-weighting showed that the folds with less number of proteins increase the sensitivity 
after re-weighting and vice-versa. However, the trend is different between simple and hierarchical re-weighting. For example, although all folds in all-$\alpha$ class have the same weight with hierarchical re-weighting  only three folds (a.3, a.39, and a.118) have similar or better sensitivity compared with simple re-weighting. 
On the other hand, sensitivity of fold a.4 drastically decreased from 49\% to 33\%. This might be due to the fact that several proteins belonging to all-$\alpha$, all-$\beta$ and $\alpha/\beta$ proteins are misclassified into $\alpha+\beta$. Further, the data presented in 
Table 
%\ref{table:sens. cross. h}
3(b) showed that folds belonging to both all-$\alpha$ and all-$\beta$ classes are misclassified into $\alpha+\beta$ class. 

\subsection*{Comparison among different re-weighting procedures}
 The results presented in Tables 2 and 3 showed that the sensitivity of recognizing protein folds differs significantly between different prediction methods (without, simple and hierarchical re-weighting). Hence, it would be difficult to choose the best method for fold recognition. However, it may be selected based on the interest of the users, whether the prediction can be done for proteins that are within a specific structural class or whole dataset and/or obtaining the accuracy of each fold or overall.

Usually, training and test sets of data are obtained from sequence and structure databases and are culled with sequence identity. However, these datasets do not always reflect proper representatives of all proteins in different folds, e.g., protein population in each fold. Further, the proteins available in databases such as, PDB are biased with the proteins that can be solved experimentally, which may be different from the proportion of real proteins. Hence, considering these aspects would help to develop ``good'' methods for protein fold recognition in future.
 
In essence, based on the methods and datasets used in the present work, we suggest that the performance with simple re-weighting is better than that without and hierarchical re-weighting.

\subsection*{Influence of amino acid occurrence in recognizing protein folds}

The importance of amino acid occurrence is illustrated with Figure 
%\ref{fig:fold_profile}
2(a). In this figure we show the occurrence of 
the 20 types of amino acid residues in TIM barrel fold and knottins. We noticed that TIM barrel fold has eight 
alpha helices and eight beta strands and hence the occurrence of all the residues except Cys is higher than 
that of knottins. Knottin is a small protein and hence it has lower occurrence of all the residues and due to 
the importance of Cys it has more number of Cys residues than TIM barrel fold. 
In Figure %\ref{fig:fold_profile}
2(b), we have shown the distribution of residues in ``amino acid occurrence'' space.
It is clearly seen that the two folds are separated well in this space.
We observed similar results about the variation of amino acid occurrences among different folds in our data set.

In addition, we have tested the performance of the method using amino acid
composition (i.e., amino acid occurrence/total number of residues) in each protein.
We noticed that the overall sensitivity without re-weighting decreased to 32\%
indicating the importance of amino acid occurrence (un-normalized composition)
in each fold (Table 
%\ref{table:sens}
1). Similar tendency is also observed for
discriminating $\beta$-barrel membrane proteins\cite{Gromiha06}. Hence, we suggest to use un-normalized composition for better prediction results. In fact, the normalization of amino acid composition produced the problem of co-linearity, i.e., diversity of vectors is not sufficient compared with the number of proteins.

\subsection*{Comparison with other methods}

We have compared the performance of our method with other related works in the literature.
Ding and Dubchak\cite{Ding} introduced a combined method for predicting the
folding type of a protein. They have used six parameters, amino acid
composition, secondary structure, hydrophobicity, van der Waals volume,
polarity and polarizability as attributes, and neural networks and support
vector machines for recognition. The features have been combined with the
number of votes in each method. They reported the sensitivity of 56\% in a
test set of 384 proteins and 10-fold cross validation sensitivity of 45\% in a
training set of 311 proteins from 27 folding types. We have used the same
dataset of 311 proteins and assessed the performance of our method. We
observed that our method could predict with the leave-one-out cross validation
accuracy of 44\%, which is similar to that (45\%) reported in Ding and
Dubchak\cite{Ding}.

In addition, we have selected the proteins from the folds that are common in
both the studies and tested the performance of our method (trained with our dataset of 1612 proteins) in predicting
the folding types of the proteins used in Ding and Dubchak\cite{Ding}.
The results are
presented in Table 
%\ref{table:Ding_Dubchak}
4. Interestingly, our method could predict the proteins
belonging to cytochrome C fold to the sensitivity of 0.94. Further, our
method with re-weighting could correctly identify the folding types with the sensitivity of
more than 0.3 in 13 among the 19 considered folds. The average sensitivity
is similar to the one that we reported with the dataset of 1612 proteins.
Although our method is optimized with different dataset it has the power to
predict the folding type of independent dataset of proteins with similar
sensitivity.

Further, there are several advantages in our method: (i) only one feature,
amino acid occurrence is sufficient for prediction rather than six features.
The comparison of results obtained with only one feature showed that the
performance of our method (45\%) is significantly better than that of Ding and
Dubchak\cite{Ding} reported with amino acid composition (20-49\%), (ii) voting procedure is not necessary and our method can be directly used for multi-fold classifications, (iii) our method uses LDA, which requires significantly less computational power compared with SVM. In SVM one has to diagonalize the matrix with the size of
(protein number) $\times$ (protein number); on the other hand, LDA
requires only diagonalization of 20 (the number of kinds of amino acid
residues) $\times$ 20 matrix independent of
number of proteins and (iv) although they have reported the dependency of fold specific sensitivities upon number of proteins in each fold, it is difficult to compensate this effect without modifying the complicated voting systems; our method has freedom to compensate it
as shown in the previous sections.

Recently, Shen and Chou\cite{Shen06} reported better sensitivity for the same
data set of Ding and Dubchak\cite{Ding}. However, the results are biased with training 
set of data. We have evaluated the sensitivity of identifying proteins belonging to 
the folds, four helical up and down bundle (a.24) and EF hand-like (a.39) and we observed 
that the sensitivity is 30.5 \% and 24 \%, respectively. Our predicted accuracies (39 \% and 
44 \%) are better than that of Shen and Chou\cite{Shen06}.

\subsection*{Fold recognition on the web}

We have developed a web server for recognizing protein folds from amino acid sequence. It takes 
the amino acid sequence as input and displays the folding type in the output. Further, the server 
has the feasibility of selecting the method, with, without and hierarchical re-weighting. It is 
freely available at http://granular.com/PROLDA/ \cite{PROLDA}.

%%%%%%%%%%%%%%%%%%%%%%
\section*{Conclusions}

In this paper, we have proposed a simple method for 
protein fold  prediction, where both the number of folds  and
the number of proteins are extensive. Interestingly, the simplest method
is the best method for the truly complicated problems.
Although complicated methods have several possibilities for tuning they generate over fitting to the data set. Further, the simple method proposed 
in this work is better than or comparable to other complicated methods, such as, 
amino acid principal principal component analysis, neural networks and support vector machines proposed in the literature for 
fold recognition. In addition, our method has several advantages including the less 
computational time and classifying the folds at a single run rather than pairwise 
comparisons. We have developed a web server\cite{PROLDA},
which takes the amino acid sequence as the 
input and displays the folding type in the output.

%%%%%%%%%%%%%%%%%%
\section*{Methods}

\subsection*{Dataset}
We have used a dataset of 1612 globular proteins belonging to 30 major folding
types obtained from SCOP database\cite{Murzin} for recognizing protein folds. This dataset has been constructed with the following criteria: (i) there should be at least 25 proteins in each fold and (ii) the sequence identity between any two proteins is not more than 25\%. 
The amino acid sequences of all the
proteins are available at \cite{PROLDA}.

  \subsection*{Linear discriminant analysis}

We have employed LDA in this work and the description of it is given below. First, we compute the amino acid occurrence of each protein,
$$
{\bf n}_i \equiv (n_{i1},n_{i2},{\ldots},n_{ij},{\ldots}  ,n_{i20}),
$$
where $n_{ij}$ is number of $j$th amino acid in $i$th protein.
Then LDA tries to maximize
$$
\eta^2 = \frac{S_B}{S_T}.
$$
$S_B$($S_T$) is the summation of squared distance between the center of mass
of all proteins and that  within fold (coordinate of each protein) along axis $z$, i.e.,
\begin{eqnarray*}
S_B & \equiv& \sum_{k=1}^{K} N_k (\bar{z} - z_k)^2 \\
S_T & \equiv& \sum_i (\bar{z} - z_i)^2, 
\end{eqnarray*}
where $K$ is the number of folds,
$N_k$ is the number of proteins belonging to $k$th fold and
$\bar{z}$ is the center  of mass along the axis $z$, and
$z_k$ is that within $k$th fold, i.e.,
\begin{eqnarray*}
\bar{z} &\equiv& \frac{1}{N} \sum_i z_i \\
z_k  &\equiv& \frac{1}{N_k} \sum_{i'=1}^{N_k} z_{i'},
\end{eqnarray*}
where $i'$ is the $i'$th protein within the $k$th fold.
$z_i$ is the linear combination of $n_{ij}$ with the set of
coefficients ${\bf a} \equiv (a_0, a_1,{\ldots}, a_j,{\ldots} a_{20})$,
$$
z_i \equiv a_0 + \sum_j a_j n_{ij}
$$
Hence, LDA tries to find ${\bf a}$ which maximizes $\eta^2$.
In total, we can get 20 kinds of $z_i$s which are orthogonal to each other, and discrimination is done based on these $z_i$s.
In addition one can introduce weights $W_k$ for each group using the equation: 
\begin{equation}
S_B  \equiv \sum_{k=1}^{K} W_k (\bar{z} - z_k)^2 \label{eq:SB}
\end{equation}

Although we have used lda module in MASS library of R\cite{R}, computational time is less than few seconds using Intel Pentium M processor (1.10GHz) and 1 GB memory.

%%%%%%%%%%%%%%%%%%%%%%%%%%%%%%%%
\section*{Authors contributions}
YhT coded the program, carried out most of the calculations and constructed
the prediction server.  MMG directly supervised the work and provided the
dataset of amino acid sequences. All authors contributed in the preparation
of the manuscript, read and approved it. None of the authors have any
competing financial or other interests in relation to this work.

%%%%%%%%%%%%%%%%%%%%%%%%%%%

%%  Bmc_article.bst  will be used to                       %%
%%  create a .BBL file for submission, which includes      %%
%%  XML structured for BMC.                                %%

{\ifthenelse{\boolean{publ}}{\footnotesize}{\small}
 \bibliographystyle{bmc_article}  % Style BST file
  \bibliography{fold_ZZJan07} }     % Bibliography file (usually '*.bib' ) 


\begin{thebibliography}{10}
\providecommand{\url}[1]{[#1]}
\providecommand{\urlprefix}{}

\bibitem{Klein}
Klein P: \textbf{Prediction of protein structural class by discriminant
  analysis}. \emph{Biochim. Biophys. Acta.} 1986, \textbf{874}:205--215.

\bibitem{Chou_2}
Chou KC, Zhang CT: \textbf{Diagrammatization of codon usage in 339 human
  immunodeficiency virus proteins and its biological implication}. \emph{AIDS
  Res. Hum. Retroviruses} 1992, \textbf{8}:1967--1976.

\bibitem{Gromiha95}
Gromiha MM, Ponnuswamy PK: \textbf{Prediction of protein secondary structures
  from their hydrophobic characteristics}. \emph{Int. J. Pept. Protein Res.}
  1995, \textbf{45}:225--240.

\bibitem{Bu}
Bu WS, Feng ZP, Zhang Z, Zhang CT: \textbf{Prediction of protein (domain)
  structural classes based on amino-acid index,}. \emph{Eur. J. Biochem.} 1999,
  \textbf{266}:1043--1049.

\bibitem{Wang}
Wang ZZ, Yuan Z: \textbf{How good is prediction of protein structural class by
  the component-coupled method?} \emph{Proteins} 2000, \textbf{38}:165 -- 175.

\bibitem{Kumarevel}
S KT, M GM, N PM: \textbf{Structural class prediction: an application of
  residue distribution along the sequence}. \emph{Biophys Chem.} 2000,
  \textbf{88}:81--101.

\bibitem{Cai}
Cai YD, Chou KC: \textbf{Predicting subcellular localization of proteins in a
  hybridization space}. \emph{Bioinformatics} 2004, \textbf{20}:1151--1156.

\bibitem{Shen05}
Shen HB, Yang J, Liu XJ, Chou KC: \textbf{Using supervised fuzzy clustering to
  predict protein structural classes}. \emph{Biochem. Biophys. Res. Commun.}
  2005, \textbf{334}:577--581.

\bibitem{Du}
Du QS, Jiang ZQ, He WZ, Li DP, Chou KC: \textbf{Amino Acid Principal Component
  Analysis (AAPCA) and its application in protein structural class prediction}.
  \emph{J. Bio. Str. Dyn.} 2006, \textbf{23}:635--640.

\bibitem{Shi}
Shi J, Blundell TL, Mizuguchi K: \textbf{FUGUE: sequence-structure homology
  recognition using environment-specific substitution tables and
  structure-dependent gap penalties,}. \emph{J Mol. Biol.} 2001,
  \textbf{310}:243--257.

\bibitem{Zhou}
Zhou H, Y Z: \textbf{Fold recognition by combining sequence profiles derived
  from evolution and from depth-dependent structural alignment of fragments}.
  \emph{Proteins} 2005, \textbf{58}:321--328.

\bibitem{Cheng}
Cheng J, Baldi P: \textbf{A machine learning informationretrieval approach to
  protein fold recognition}. \emph{Bioinformatics} 2006, \textbf{22}:1456--63.

\bibitem{Gromiha05}
Gromiha MM, Suwa M: \textbf{A Simple statistical method for discriminating
  outer membrane proteins with better accuracy}. \emph{Bioinformatics} 2005,
  \textbf{21}:961--968.

\bibitem{Hirokawa}
Hirokawa T, Boon-Chieng S, Mitaku S: \textbf{SOSUI: classification and
  secondary structure prediction system for membrane proteins}.
  \emph{Bioinformatics} 1998, \textbf{14}:378--379.

\bibitem{Chou}
Chou KC: \textbf{Prediction of protein structural classes and subcellular
  locations}. \emph{Curr. Protein Pept. Sci.} 2000, \textbf{1}:171--208.

\bibitem{Gromiha06}
Gromiha MM, Selvaraj S, Thangakani AM: \textbf{A Statistical method for
  predicting protein unfolding rates from amino acid sequence}. \emph{J. Chem.
  Inf. Model} 2006, \textbf{46}:1503--1508.

\bibitem{Gromiha99}
Gromiha MM, Oobatake M, Kono H, Uedaira H, Sarai A: \textbf{Relationship
  between amino acid properties and protein stability: Buried Mutations}.
  \emph{J. Protein Chem.} 1999, \textbf{18}:565--578.

\bibitem{Ding}
Ding HQD, Dubchak I: \textbf{Multi-class protein fold recognition using support
  vector machines and neural networks,}. \emph{Bioinformatics} 2001,
  \textbf{17}:349--358.

\bibitem{Shen06}
Shen HB, Chou KC: \textbf{Ensemble classifier for protein fold pattern
  recognition}. \emph{Bioinformatics} 2006, \textbf{22}:1717--1722.

\bibitem{PROLDA}
\textbf{PROLDA} \urlprefix\url{[http://granular.com/PROLDA/]}.

\bibitem{Murzin}
Murzin AG, Brenner SE, Hubbard T, Chothia C: \textbf{SCOP: a structural
  classification of proteins database for the investigation of sequences and
  structures}. \emph{J. Mol. Biol.} 1995, \textbf{247}:536--540.

\bibitem{R}
R Development Core Team: \emph{R: A language and environment for statistical
  computing}. http://www.R-project.org 2005.

\end{thebibliography}

\newcommand{\BMCxmlcomment}[1]{}

\BMCxmlcomment{

<refgrp>

<bibl id="B1">
  <title><p>Prediction of protein structural class by discriminant
  analysis</p></title>
  <aug>
    <au><snm>Klein</snm><fnm>P</fnm></au>
  </aug>
  <source>Biochim. Biophys. Acta.</source>
  <pubdate>1986</pubdate>
  <volume>874</volume>
  <fpage>205</fpage>
  <lpage>215</lpage>
</bibl>

<bibl id="B2">
  <title><p>Diagrammatization of codon usage in 339 human immunodeficiency
  virus proteins and its biological implication</p></title>
  <aug>
    <au><snm>Chou</snm><fnm>K C</fnm></au>
    <au><snm>Zhang</snm><fnm>C T</fnm></au>
  </aug>
  <source>AIDS Res. Hum. Retroviruses</source>
  <pubdate>1992</pubdate>
  <volume>8</volume>
  <fpage>1967</fpage>
  <lpage>1976</lpage>
</bibl>

<bibl id="B3">
  <title><p>Prediction of protein secondary structures from their hydrophobic
  characteristics</p></title>
  <aug>
    <au><snm>Gromiha</snm><fnm>M M</fnm></au>
    <au><snm>Ponnuswamy</snm><fnm>P K</fnm></au>
  </aug>
  <source>Int. J. Pept. Protein Res.</source>
  <pubdate>1995</pubdate>
  <volume>45</volume>
  <fpage>225</fpage>
  <lpage>240</lpage>
</bibl>

<bibl id="B4">
  <title><p>Prediction of protein (domain) structural classes based on
  amino-acid index,</p></title>
  <aug>
    <au><snm>Bu</snm><fnm>W S</fnm></au>
    <au><snm>Feng</snm><fnm>Z P</fnm></au>
    <au><snm>Zhang</snm><fnm>Z</fnm></au>
    <au><snm>Zhang</snm><fnm>C T</fnm></au>
  </aug>
  <source>Eur. J. Biochem.</source>
  <pubdate>1999</pubdate>
  <volume>266</volume>
  <fpage>1043</fpage>
  <lpage>1049</lpage>
</bibl>

<bibl id="B5">
  <title><p>How good is prediction of protein structural class by the
  component-coupled method?</p></title>
  <aug>
    <au><snm>Wang</snm><fnm>Z Z</fnm></au>
    <au><snm>Yuan</snm><fnm>Z</fnm></au>
  </aug>
  <source>Proteins</source>
  <pubdate>2000</pubdate>
  <volume>38</volume>
  <fpage>165</fpage>
  <lpage>175</lpage>
</bibl>

<bibl id="B6">
  <title><p>Structural class prediction: an application of residue distribution
  along the sequence</p></title>
  <aug>
    <au><snm>S</snm><fnm>KT</fnm></au>
    <au><snm>M</snm><fnm>GM</fnm></au>
    <au><snm>N</snm><fnm>PM</fnm></au>
  </aug>
  <source>Biophys Chem.</source>
  <pubdate>2000</pubdate>
  <volume>88</volume>
  <fpage>81</fpage>
  <lpage>101</lpage>
</bibl>

<bibl id="B7">
  <title><p>Predicting subcellular localization of proteins in a hybridization
  space</p></title>
  <aug>
    <au><snm>Cai</snm><fnm>Y D</fnm></au>
    <au><snm>Chou</snm><fnm>K C</fnm></au>
  </aug>
  <source>Bioinformatics</source>
  <pubdate>2004</pubdate>
  <volume>20</volume>
  <fpage>1151</fpage>
  <lpage>1156</lpage>
</bibl>

<bibl id="B8">
  <title><p>Using supervised fuzzy clustering to predict protein structural
  classes</p></title>
  <aug>
    <au><snm>Shen</snm><fnm>H B</fnm></au>
    <au><snm>Yang</snm><fnm>J</fnm></au>
    <au><snm>Liu</snm><fnm>X J</fnm></au>
    <au><snm>Chou</snm><fnm>K C</fnm></au>
  </aug>
  <source>Biochem. Biophys. Res. Commun.</source>
  <pubdate>2005</pubdate>
  <volume>334</volume>
  <fpage>577</fpage>
  <lpage>581</lpage>
</bibl>

<bibl id="B9">
  <title><p>Amino Acid Principal Component Analysis (AAPCA) and its application
  in protein structural class prediction</p></title>
  <aug>
    <au><snm>Du</snm><fnm>Q S</fnm></au>
    <au><snm>Jiang</snm><fnm>Z Q</fnm></au>
    <au><snm>He</snm><fnm>W Z</fnm></au>
    <au><snm>Li</snm><fnm>D P</fnm></au>
    <au><snm>Chou</snm><fnm>K C</fnm></au>
  </aug>
  <source>J. Bio. Str. Dyn.</source>
  <pubdate>2006</pubdate>
  <volume>23</volume>
  <fpage>635</fpage>
  <lpage>640</lpage>
</bibl>

<bibl id="B10">
  <title><p>FUGUE: sequence-structure homology recognition using
  environment-specific substitution tables and structure-dependent gap
  penalties,</p></title>
  <aug>
    <au><snm>Shi</snm><fnm>J</fnm></au>
    <au><snm>Blundell</snm><fnm>T L</fnm></au>
    <au><snm>Mizuguchi</snm><fnm>K</fnm></au>
  </aug>
  <source>J Mol. Biol.</source>
  <pubdate>2001</pubdate>
  <volume>310</volume>
  <fpage>243</fpage>
  <lpage>257</lpage>
</bibl>

<bibl id="B11">
  <title><p>Fold recognition by combining sequence profiles derived from
  evolution and from depth-dependent structural alignment of
  fragments</p></title>
  <aug>
    <au><snm>Zhou</snm><fnm>H</fnm></au>
    <au><snm>Y</snm><fnm>Z</fnm></au>
  </aug>
  <source>Proteins</source>
  <pubdate>2005</pubdate>
  <volume>58</volume>
  <fpage>321</fpage>
  <lpage>328</lpage>
</bibl>

<bibl id="B12">
  <title><p>A machine learning informationretrieval approach to protein fold
  recognition</p></title>
  <aug>
    <au><snm>Cheng</snm><fnm>J</fnm></au>
    <au><snm>Baldi</snm><fnm>P</fnm></au>
  </aug>
  <source>Bioinformatics</source>
  <pubdate>2006</pubdate>
  <volume>22</volume>
  <fpage>1456</fpage>
  <lpage>63</lpage>
</bibl>

<bibl id="B13">
  <title><p>A Simple statistical method for discriminating outer membrane
  proteins with better accuracy</p></title>
  <aug>
    <au><snm>Gromiha</snm><fnm>M M</fnm></au>
    <au><snm>Suwa</snm><fnm>M</fnm></au>
  </aug>
  <source>Bioinformatics</source>
  <pubdate>2005</pubdate>
  <volume>21</volume>
  <fpage>961</fpage>
  <lpage>968</lpage>
</bibl>

<bibl id="B14">
  <title><p>SOSUI: classification and secondary structure prediction system for
  membrane proteins</p></title>
  <aug>
    <au><snm>Hirokawa</snm><fnm>T</fnm></au>
    <au><snm>Boon Chieng</snm><fnm>S</fnm></au>
    <au><snm>Mitaku</snm><fnm>S</fnm></au>
  </aug>
  <source>Bioinformatics</source>
  <pubdate>1998</pubdate>
  <volume>14</volume>
  <fpage>378</fpage>
  <lpage>379</lpage>
</bibl>

<bibl id="B15">
  <title><p>Prediction of protein structural classes and subcellular
  locations</p></title>
  <aug>
    <au><snm>Chou</snm><fnm>K C</fnm></au>
  </aug>
  <source>Curr. Protein Pept. Sci.</source>
  <pubdate>2000</pubdate>
  <volume>1</volume>
  <fpage>171</fpage>
  <lpage>208</lpage>
</bibl>

<bibl id="B16">
  <title><p>A Statistical method for predicting protein unfolding rates from
  amino acid sequence</p></title>
  <aug>
    <au><snm>Gromiha</snm><fnm>M M</fnm></au>
    <au><snm>Selvaraj</snm><fnm>S</fnm></au>
    <au><snm>Thangakani</snm><fnm>A M</fnm></au>
  </aug>
  <source>J. Chem. Inf. Model</source>
  <pubdate>2006</pubdate>
  <volume>46</volume>
  <fpage>1503</fpage>
  <lpage>1508</lpage>
</bibl>

<bibl id="B17">
  <title><p>Relationship between amino acid properties and protein stability:
  Buried Mutations</p></title>
  <aug>
    <au><snm>Gromiha</snm><fnm>M M</fnm></au>
    <au><snm>Oobatake</snm><fnm>M</fnm></au>
    <au><snm>Kono</snm><fnm>H</fnm></au>
    <au><snm>Uedaira</snm><fnm>H</fnm></au>
    <au><snm>Sarai</snm><fnm>A</fnm></au>
  </aug>
  <source>J. Protein Chem.</source>
  <pubdate>1999</pubdate>
  <volume>18</volume>
  <fpage>565</fpage>
  <lpage>578</lpage>
</bibl>

<bibl id="B18">
  <title><p>Multi-class protein fold recognition using support vector machines
  and neural networks,</p></title>
  <aug>
    <au><snm>Ding</snm><fnm>H Q D</fnm></au>
    <au><snm>Dubchak</snm><fnm>I</fnm></au>
  </aug>
  <source>Bioinformatics</source>
  <pubdate>2001</pubdate>
  <volume>17</volume>
  <fpage>349</fpage>
  <lpage>358</lpage>
</bibl>

<bibl id="B19">
  <title><p>Ensemble classifier for protein fold pattern
  recognition</p></title>
  <aug>
    <au><snm>Shen</snm><fnm>H B</fnm></au>
    <au><snm>Chou</snm><fnm>K C</fnm></au>
  </aug>
  <source>Bioinformatics</source>
  <pubdate>2006</pubdate>
  <volume>22</volume>
  <fpage>1717</fpage>
  <lpage>1722</lpage>
</bibl>

<bibl id="B20">
  <title><p>PROLDA</p></title>
  <url>http://granular.com/PROLDA/</url>
</bibl>

<bibl id="B21">
  <title><p>SCOP: a structural classification of proteins database for the
  investigation of sequences and structures</p></title>
  <aug>
    <au><snm>Murzin</snm><fnm>A G</fnm></au>
    <au><snm>Brenner</snm><fnm>S E</fnm></au>
    <au><snm>Hubbard</snm><fnm>T</fnm></au>
    <au><snm>Chothia</snm><fnm>C</fnm></au>
  </aug>
  <source>J. Mol. Biol.</source>
  <pubdate>1995</pubdate>
  <volume>247</volume>
  <fpage>536</fpage>
  <lpage>540</lpage>
</bibl>

<bibl id="B22">
  <title><p>R: A language and environment for statistical computing</p></title>
  <aug><au><cnm>R Development Core Team</cnm></au></aug>
  <publisher>http://www.R-project.org</publisher>
  <pubdate>2005</pubdate>
</bibl>

</refgrp>
} % end of \BMCxmlcomment

%%%%%%%%%%%

\ifthenelse{\boolean{publ}}{\end{multicols}}{}

%%%%%%%%%%%%%%%%%%%%%%%%%%%%%%%%%%%
%%                               %%
%% Figures                       %%
%%                               %%
%% NB: this is for captions and  %%
%% Titles. All graphics must be  %%
%% submitted separately and NOT  %%
%% included in the Tex document  %%
%%                               %%
%%%%%%%%%%%%%%%%%%%%%%%%%%%%%%%%%%%

%%
%% Do not use \listoffigures as most will included as separate files

\section*{Figures}

\subsection*{Figure 1 - Prediction versus experiment}

\includegraphics{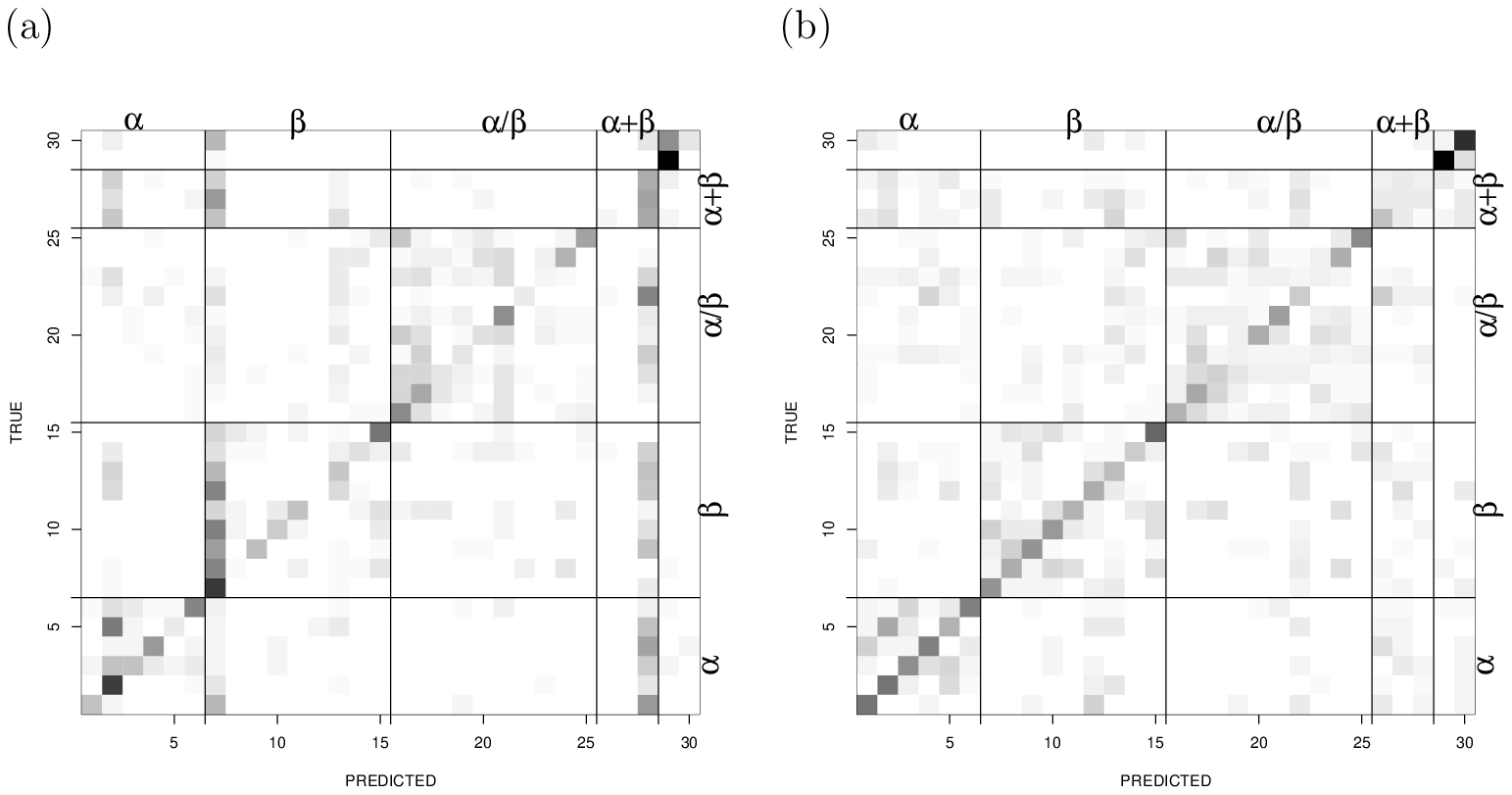}

Comparison between predicted and experimental folds in 1612 proteins. The diagonal elements show the correctly predicted proteins. Dark block indicates the presence of more number of proteins and solid line indicates boundary between five classes as shown in
Table 
%\ref{table:sens. set 1}
2, i.e., all-$\alpha$, all-$\beta$,
$\alpha/\beta$, and $\alpha+\beta$ and small proteins.
(a)without reweighing. (b) with reweighing.\label{fig:matrix}

\subsection*{Figure 2 - Amino acid occurrence}

\includegraphics{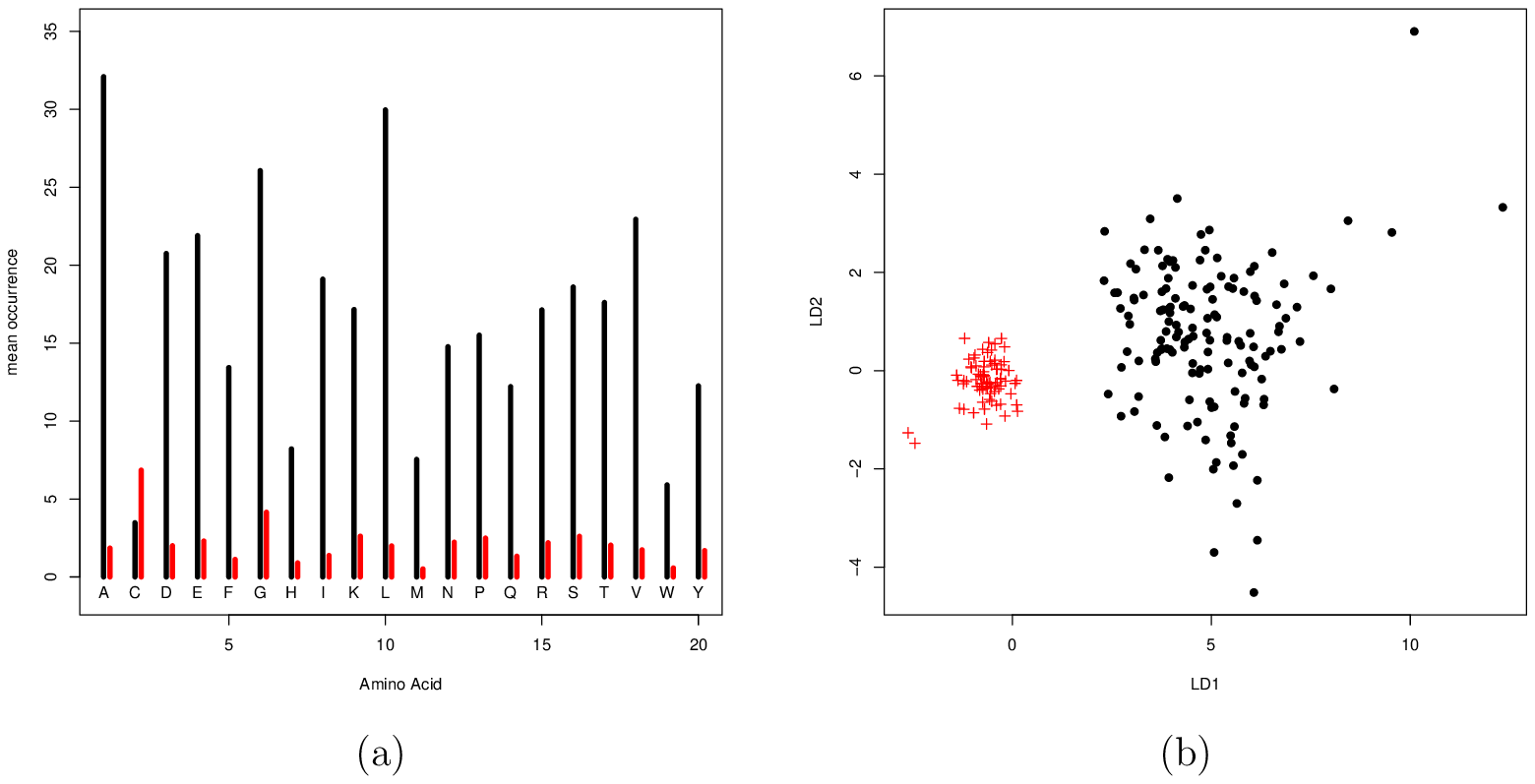}

(a)Comparison between mean amino acid occurrences of the most distant
pair of folds, TIM barrel (black) and  knottins (red). 
(b) Distribution of these two folds over the first two discriminant
functions with re-weighting.
\label{fig:fold_profile}

%%%%%%%%%%%%%%%%%%%%%%%%%%%%%%%%%%%
%%                               %%
%% Tables                        %%
%%                               %%
%%%%%%%%%%%%%%%%%%%%%%%%%%%%%%%%%%%

%% Use of \listoftables is discouraged.
%%
\section*{Tables}
  \subsection*{Table 1 - Role of re-weighting}
{ Leave-one-out cross validation results for two types of sensitivities.\label{table:sens}}
\par \mbox{}
    \par
    \mbox{
\begin{tabular}{ccccc}
\hline
&  \multicolumn{2}{c}{with re-weighting}         &
\multicolumn{2}{c}{without re-weighting} 
\\\cline{2-5}
	       &  over all & fold average & over all & fold average \\\hline
Occurrence          &  0.33     & 0.32   &   0.37 & 0.28 \\
Composition       &  0.26    &  0.27   &   0.32 & 0.24 \\
\hline
\end{tabular}
      }
  \subsection*{Table 2 - Sensitivity of fold recognition}
Leave-one-out cross validation sensitivity in each fold.
\par \mbox{}
    \par
    \mbox{\tiny
\begin{tabular}{rclcccc}
\hline
 & & &  & \multicolumn{3}{c}{Sensitivity(\%)}   \\\cline{5-7}
ID &  Fold & Fold Description& Number & \multicolumn{1}{l}{without re-}   &
\multicolumn{1}{l}{with re-}  &    \\
 & &  &  &weighting & weighting &  hierarchical   \\\hline
\multicolumn{7}{c}{ } \\
\multicolumn{3}{c}{all-$\alpha$} & \multicolumn{4}{c}{ } \\
1 &   a.3  & Cytochrome C &   25 & 24 & 48 & 48 \\
 2 &   a.4  & DNA/RNA binding 3-helical bundle &  103 & 72 & 49 & 33 \\
 3 &   a.24 & Four helical up and down bundle &   26 & 23 & 39 & 34 \\
 4 &   a.39 & EF hand-like fold &   25 & 40 & 44 & 44 \\
 5 &   a.60 & SAMdomain-like &   26 &  7 & 27 & 19 \\
 6 &   a.118& $\alpha$-$\alpha$ superhelix &   47 & 46 & 45 & 46 \\
\multicolumn{7}{c}{ }\\
\multicolumn{3}{c}{all-$\beta$} & \multicolumn{4}{c}{ }		\\
 7 &   b.1  & Immunoglobulin-like $\beta$-sandwich &  173 & 76 & 38 & 13  \\ 
 8 &   b.2  & Common fold of diphtheria toxin/transcription factors/cytochrome f&   28 &  3 & 29 & 28  \\
 9 &   b.6  & Cupredoxin-like &   30 & 26 & 37 & 30 \\
10 &  b.18  & Galactose-binding domain-like &   25 & 20 & 36 & 36 \\
11 &  b.29  & Concanavalin A-like lectins/glucanases &   26 & 23 & 27 & 26 \\
12 &  b.34  & SH3-like barrel &   42 &  0 & 28 &  7 \\
13 &  b.40  & OB-fold &   78 & 21 & 24 &  7 \\
14 &  b.82  & Double-stranded a-helix &   34 & 11 & 18 & 20 \\
15 &  b.121 & Nucleoplasmin-like&   42 & 52 & 52 & 50 \\
\multicolumn{7}{c}{ }\\
\multicolumn{3}{c}{$\alpha/\beta$} & \multicolumn{4}{c}{ }  \\
16 &  c.1   & TIM barrel &  145 & 44 & 26 & 25 \\
17 &  c.2   & NAD(P)-binding Rossmann-fold domains&   77 & 33 & 31 & 29 \\
18 &  c.3   & FAD/NAD(P)-binding domain &   31 &  9 & 16 & 16 \\
19 & c.23   & Flavodoxin-like &   55 & 10 &  5 &  3 \\
20 & c.26   & Adenine nucleotide a hydrolase-like &   34 & 11 & 29 & 26 \\
21 & c.37   & P-loop containing nucleoside triphosphate hydrolases &   95 & 43 & 33 & 32 \\
22 & c.47   & Thioredoxin fold &   32 &  9 & 18 &  9 \\
23 & c.55   & Ribonuclease H-like motif &   49 &  4 &  6 &  4 \\
24 & c.66   & S-adenosyl-L-methionine-dependent methyltransferases &   34 & 29 & 29 & 29 \\
25 & c.69   & $\alpha$/$\beta$-Hydrolases &   37 & 35 & 40 & 40 \\
\multicolumn{7}{c}{ }\\
\multicolumn{3}{c}{$\alpha+\beta$} & \multicolumn{4}{c}{ } \\
26 & d.15   & $\beta$-Grasp, ubiquitin-like &   42 &  4 & 21 & 35 \\
27 & d.17   & Cystatin-like &   25 &  0 &  8 & 20 \\
28 & d.58   & Ferredoxin-like &  118 & 32 &  7 & 16 \\
\multicolumn{7}{c}{ }\\
\multicolumn{3}{c}{small} & \multicolumn{4}{c}{ } \\
29 & g.3    & Knottins &   80 & 97 & 88 & 88 \\
30 & g.41   & Rubredoxin-like &   28 & 10 & 71 & 85 \\\hline
\end{tabular}
}
\subsection*{Table 3 - Sensitivity among structural classes}
Leave-One-out results of true vs predicted structural classes 
(a) with simple and (b) hierarchical re-weighting.
\label{table:sens. cross. h}
\par \mbox{}
    \par
    \mbox{
\begin{tabular}{l}
(a) simple reweighting\\
\begin{tabular}{crrrrrr}
\hline
 & \multicolumn{5}{c}{predicted} &  \\\cline{2-6}
True  &  all-$\alpha$  &  all-$\beta$ & $\alpha/\beta$  &   $\alpha+\beta$  &
small &total \\\hline
all-$\alpha$    &       178 &  32  &  13 &  20 &    9 & 252\\
all-$\beta$     &        41 & 320  &  56 &  42 &   19 & 478 \\
$\alpha/\beta$  &        61 &  89  & 413 &  25 &    1 & 589 \\
$\alpha+\beta$  &        54 &  34  &  34 &  44 &   19 & 185 \\
small           &         3 &   3  &   0 &   1 &  101 & 108\\\hline
\end{tabular} \\
\\
 (b) hierarchical reweighting\\
\begin{tabular}{crrrrrr}
\hline
 & \multicolumn{5}{c}{predicted} &  \\\cline{2-6}
True  &  all-$\alpha$  &  all-$\beta$ & $\alpha/\beta$  &   $\alpha+\beta$  &
small & total\\\hline
all-$\alpha$    &  144      &   4  &   3 &   84 &  17 & 252 \\
all-$\beta$     &   20      & 180  &  27 &  203 &  48 & 478 \\
$\alpha/\beta$  &   62      &  61  & 372 &   90 &   4 & 589 \\
$\alpha+\beta$  &   30      &   8  &  12 &  109 &  26 & 185 \\
small           &    0      &   1  &   0 &    2 & 105 & 108 \\\hline
\end{tabular}
\end{tabular}
}

\subsection*{Table 4 - Sensitivity with independent dataset}
Predictive ability of our method to the independent dataset of proteins used
in Ding and Dubchak\cite{Ding}. \label{table:Ding_Dubchak}
\par \mbox{}
    \par
    \mbox{
\begin{tabular}{lccc}
\hline
 \multicolumn{2}{r}{} &  \multicolumn{2}{c}{Sensitivity (\%)} \\\cline{3-4} 
Fold Description & Number &  without re- &with re- \\
\multicolumn{2}{c}{} &  weighting     & weighting \\\hline
                                         Cytochrome C &16 &  56  & 94  \\
                     DNA/RNA binding 3-helical bundle &32 &  75  & 56  \\
                      Four helical up and down bundle &15 &  33  & 33  \\
                                    EF hand-like fold &15 &  53  & 53  \\
                      Immunoglobulin-like $\beta$-sandwich  &74 &  66  & 31  \\
                                      Cupredoxin-like &21 &  29  & 38  \\
        Concanavalin A-like lectins/glucanases &13 &  38  & 38  \\
                                      SH3-like barrel &16 &   0  & 50  \\
                                              OB-fold &32 &  16  & 28  \\
                                       TIM barrel &77 &  40  & 25  \\
                          FAD/NAD: (P)-binding domain &23 &  22  & 30  \\
                                      Flavodoxin-like &24 &  8   & 13 \\
               NAD: (P)-binding Rossmann-fold domains &40 &  40  & 35  \\
 P-loop containing nucleoside triphosphate hydrolases &22 &  23  & 18  \\
                                     Thioredoxin fold &17 &  18  & 35  \\
                            Ribonuclease H-like motif &22 &   5  & 18  \\
                          $\alpha$/$\beta$-Hydrolases &18 &  33  & 39  \\
                        $\beta$-Grasp, ubiquitin-like &15 &   0  & 33  \\
                                      Ferredoxin-like &40 &  23  &  3 \\\hline
                                        over all      &532&  36  & 32  \\
                                        fold average  &   &  30  & 35 \\\hline
\end{tabular}
}

%%%%%%%%%%%%%%%%%%%%%%%%%%%%%%%%%%%
%%                               %%
%% Additional Files              %%
%%                               %%
%%%%%%%%%%%%%%%%%%%%%%%%%%%%%%%%%%%

\end{bmcformat}
\end{document}